\documentclass[preprint2]{aastex}
\usepackage{aastexug} 

\def\lesssim{\mathrel{\hbox{\rlap{\hbox{\lower4pt\hbox{$\sim$}}}\hbox{$<$}}}}
\def\gtrsim{\mathrel{\hbox{\rlap{\hbox{\lower4pt\hbox{$\sim$}}}\hbox{$>$}}}}

\def\plotfiddle#1#2#3#4#5#6#7{\centering \leavevmode
\vbox to#2{\rule{0pt}{#2}}
\includegraphics{#1}}

\begin{document}

\title{Comparison of the Two Followup Observation Strategies \\
       for Gravitational Microlensing Planet Searches}

\author{Cheongho Han \& Yong-Gi Kim}
\affil{Department of Astronomy \& Space Science, \\
       Chungbuk National University, Chongju, Korea 361-763}
\email{cheongho,ykkim@astro.chungbuk.ac.kr}
\vskip-2cm

\vspace{-\abovedisplayskip} 
\begin{abstract}
There are two different strategies of followup observations for the detection 
of planets by using microlensing.  One is detecting the light curve anomalies 
affected by the planetary caustic from continuous monitoring of all events 
detected by microlensing survey programs (type I strategy) and the other is 
detecting anomalies near the peak amplification affected by the central caustic
from intensive monitoring of high amplification events (type II strategy).  It 
was shown by Griest \& Safizadeh that the type II strategy yields high planet 
detection efficiency per event.  However, it is not known the planet detection 
rate by this strategy can make up a substantial fraction of the total rate.  
In this paper, we estimate the relative planet detection rates expected under 
the two followup observation strategies.  From this estimation, we find that 
the rate under the type II strategy is substantial and will comprise $\sim 1/4$
-- 1/2 of the total rate.  We also find that compared to the type I strategy 
the type II strategy is more efficient in detecting planets located outside of 
the lensing zone.  We determine the optimal monitoring frequency of the type 
II strategy to be $\sim 20$ times/night, which can be easily achieved by the 
current microlensing followup programs even with a single telescope.
\end{abstract}


\clearpage


\section{Introduction}

Experiments to detect microlensing events by monitoring millions of stars 
located in the Magellanic Clouds and the Galactic bulge have been and are 
being carried out by several groups \citep{alcock93, aubourg93, udalski93, 
alard97, abe97}.  With their efforts, the total number of candidate events 
now reaches up to a thousand (P.\ Popowski 1999, private communication).

Although the primary goal for these experiments was to investigate the nature 
of Galactic dark matter, it turns out to be that microlensing can be very 
useful in various other fields of astronomy.  One of the important applications
of microlensing is the detection of extra-solar planets.  Planet detection by 
using microlensing is possible because the event caused by a lens system with 
a planet can exhibit detectable anomalies in the light curve when the source 
passes close to the lens caustics \citep{mao91, gould92, bolatto94}.  For this 
lens system, there are 2 or 3 disconnected sets of caustics depending on the 
projected separation between the planet and the primary lens (planet-lens 
separation).  Among them, one is located very close to the primary lens, 
`central caustic', and the other(s) is (are) located relatively away from the 
primary lens, `planetary caustics' \citep{griest98}.  Accordingly, there exist 
two different types of anomalies in the light curves; one affected by the 
planetary caustic(s), `type I anomaly', and the other affected by the central 
caustic, `type II anomaly' \citep{covone00}.  Due to the characteristics of 
the central caustic, type II anomalies occur near the peak of the light curves 
of high amplification events.

Compared to the frequency of type I anomalies, type II anomalies occur with a 
relatively low frequency due to the smaller size of the central caustic than 
the corresponding planetary caustic.  However, the efficiency of detecting
type II anomalies can be high because intensive monitoring of events is 
possible due to the predictable time of anomalies \citep{griest98}.  We call 
the planet search strategy of intensive monitoring near the peak of high 
amplification events as the `type II' strategy, while the strategy of 
continuous monitoring for all events detected from lensing survey programs as 
the `type I' strategy.  Then a naturally arising questions is whether the 
planet detection rate (not the detection efficiency per event) under the type 
II strategy can make up a substantial fraction of the total rate.  In this 
paper, we answer to this question by estimating the relative detection rates 
under the two strategies of planet search followup observations.

\section{Types of Planet-induced Anomalies}

The lens system with a planet is described by the formalism of the binary 
lens system with a very low-mass companion.  When lengths are normalized 
to the combined angular Einstein ring radius, which is equivalent to the 
angular Einstein ring radius of a single lens with a mass equal to the total 
mass of the binary, the lens equation in complex notations for a binary lens 
system is represented by
\begin{equation}
\zeta = z + {m_{1} \over \bar{z}_{1}-\bar{z}} 
+ {m_{2} \over \bar{z}_{2}-\bar{z}},
\end{equation}
where $m_1$ and $m_2$ are the mass fractions of individual lenses (and thus 
$m_1+m_2=1$), $z_1$ and $z_2$ are the positions of the lenses, $\zeta=\xi 
+ i\eta$ and $z=x+iy$ are the positions of the source and images, and 
$\bar{z}$ denotes the complex conjugate of $z$ \citep{witt90}.  The combined 
angular Einstein ring radius is related to the physical parameters of the 
lens system by
\begin{equation}
\theta_{\rm E} = \left( { 4GM \over c^2 } {D_{\rm ls}\over 
            D_{\rm os}D_{\rm ol}}\right)^{1/2},
\end{equation}
where $M$ is the total mass of the binary lens system and $D_{\rm ol}$, 
$D_{\rm ls}$, and $D_{\rm os}$ represent the separations between the observer, 
lens, and source star, respectively.  The amplification of each image is given 
by the Jacobian of the transformation (1) evaluated at the image position, 
i.e.\
\begin{equation}
A_i = \left({1\over \vert {\rm det}\ J\vert} \right)_{z=z_i};
\qquad {\rm det}\ J = 1-{\partial\zeta\over\partial\bar{z}}
{\overline{\partial\zeta}\over\partial\bar{z}}.
\end{equation}
Then the total amplification of a binary lens event is given by the sum of the 
amplifications of the individual images, i.e.\ $A=\sum_i A_i$.  The set of 
source positions with infinite amplifications, i.e.\ ${\rm det}\ J=0$, form 
closed curves called caustics.  As a result, whenever a source passes very 
close to the caustic, the resulting light curve deviates significantly from 
the standard \citet{paczynski86} curve.

The caustic structure of the lens system with a planet varies depending on the 
planet-lens separation $d$ normalized by $\theta_{\rm E}$.  A lens system with 
a wide planet ($d > 1.0$) forms 2 disconnected sets of caustics.  One caustic 
is located very close to the center of mass (and thus referred as the central 
caustic), and the other planetary caustic is located away from the center of 
mass on the planet-side (with respect to the center of mass) axis connecting 
the primary lens and the planet.  A system with a close planet ($d < 1.0$) 
also has a single central caustic, but has two planetary caustics, which are 
located on the opposite side of the planet with respect to the center of mass 
and not on the planet-lens axis.  The sizes of both the central and planetary 
caustics are maximized when the planet is located in the lensing zone of 
$0.618\leq d\leq 1.618$ \citep{griest98}, and decrease as the planet-lens 
separation becomes either smaller or larger.  However, regardless of the 
separation the central caustic of a lens system is always smaller than the 
corresponding planetary caustic(s).

\begin{figure}[t]
\plotfiddle{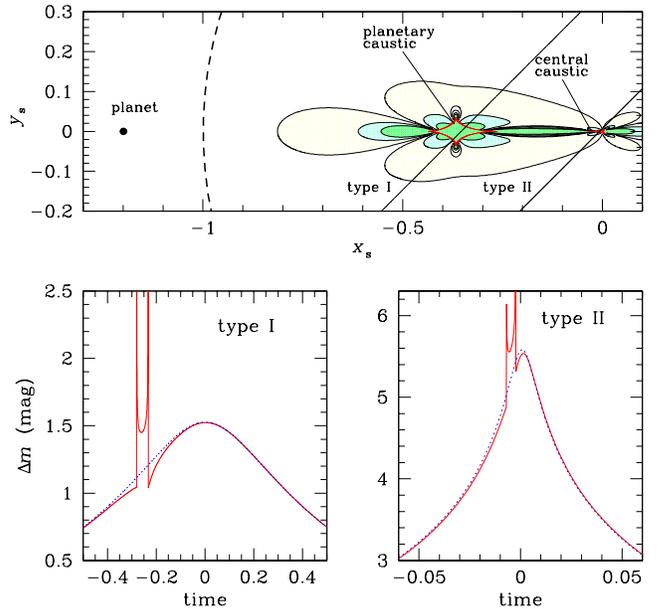}{0.0cm}{0}{43}{43}{-135}{-290}
\vskip7.6cm
\caption{The central and planetary caustics of an example lens system with 
a planet and the corresponding type I and II anomalies in the resulting 
light curves.  Presented in upper panel is the contour map of the fractional 
amplification excess $\epsilon$.  The contours around the individual types 
of caustic (closed figures drawn by the thick solid line) are drawn at the 
levels of $\epsilon = 1\%$, 5\%, and 10\%.  The example lens system has a 
mass ratio and a planet-lens separation of $q=10^{-3}$.  The lower panels 
show the light curves resulting from the two source trajectories marked by 
straight lines in the upper panel.  The two curves in each panel represent 
those expected in the presence (solid curve) and absence (dotted curve) of 
the planet, respectively.}
\end{figure}

In Figure 1, we illustrate the central and planetary caustics of an example 
lens system with a planet and the corresponding type I and II anomalies in 
the resulting light curves.  Presented in upper panel are the contour map 
of the fractional amplification excess, which is computed by
\begin{equation}
\epsilon = {\left\vert A - A_{\rm s} \right\vert \over A_{\rm s}},
\end{equation}
where $A$ and $A_{\rm s}$ represent the amplifications with and without the 
existence of the planet, respectively.  The single lens event amplification 
is related to the separation (normalized by the angular Einstein ring radius 
of the primary lens) between the source and the primary lens, $u$, by
\begin{equation}
A_{\rm s} = {u^2+ 2 \over u(u^2+4)^{1/2}}.
\end{equation}
The contours (thin solid curves) around the individual types of caustic 
(closed figures drawn by the thick solid line) are drawn at the levels of 
$\epsilon=1\%$, 5\%, and 10\% with increasing distance from the caustics.  
The example lens system has a mass ratio and a projected planet-lens 
separation of $q=10^{-3}$ (corresponding to a Jupiter-mass planet around a 
$1\ M_\odot$ star) and $d=1.2$.  The positions of the primary lens and the 
planet are chosen such that the center of mass is at the origin and all 
lengths are scaled by $\theta_{\rm E}$.  The lower panels show the light 
curves resulting from the two source trajectories marked by straight lines 
in the upper panel.  The two curves in each panel represent those expected 
in the presence (solid curve) and absence (dotted curve) of the planet.  
Times are in units of the Einstein ring radius crossing time (Einstein 
time-scale).  From the figure, one finds that the region of large type II 
deviations is smaller than the corresponding region of type I deviations.  
One also finds that the type II anomaly in the light curve occurs near 
the peak amplification.

\section{Type I versus Type II Strategies}

\begin{figure}[t]
\plotfiddle{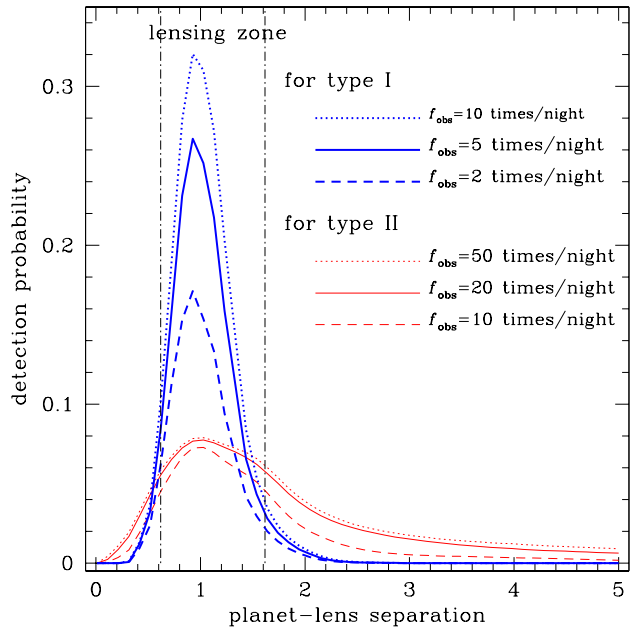}{0.0cm}{0}{45}{45}{-140}{-305}
\vskip8.0cm
\caption{Distributions of the planet detection probability expected under 
the type I and II followup observation strategies with various monitoring 
frequencies as a function of the planet-lens separation for events caused 
by a planetary system with $q=10^{-3}$.  The events are assumed to have 
$t_{\rm E}=20\ {\rm days}$.  The region enclosed by the dot-dashed lines 
represents the lensing zone with $0.618\leq d\leq 1.618$.  For the details 
about the observational conditions and detection criteria, see the text.}
\end{figure}

\begin{deluxetable}{cccc}
\tabletypesize{\small}
\tablecaption{The Fractions of Planet Detections by the Type II Strategy}
\tablewidth{0pt}
\tablehead{
\multicolumn{2}{c}{$f_{\rm obs}$} &
\multicolumn{2}{c}{fraction (\%)} \\
\multicolumn{2}{c}{(times/night)} &
\colhead{\ \ for all systems\ \ } &
\colhead{for systems with} \\
\colhead{type I} &
\colhead{type II} &
\colhead{with $d\leq 5.0$} &
\colhead{$d$ in the lensing zone} }
\startdata
2   &  10 &  45.2  &  37.6    \\
    &  20 &  52.1  &  40.5    \\
\smallskip
    &  50 &  53.6  &  41.3    \\
5   &  10 &  34.8  &  27.9    \\
    &  20 &  41.3  &  30.4    \\
\smallskip
    &  50 &  42.8  &  31.1    \\
10  &  10 &  30.9  &  23.9    \\
    &  20 &  36.4  &  26.3    \\
    &  50 &  37.8  &  26.8    \\
\enddata
\vskip-0.8cm
\tablecomments{The fractions of planet detection rate under the type II 
strategy out of the total rate for various combinations of monitoring 
frequencies of the individual strategies.  Two sets of fractions are 
presented.  The first set is for all planetary systems with planet 
separations $d\leq 5.0$ and the second set is for those with planets 
located in the lensing zone only.}
\end{deluxetable}

\begin{figure}[t]
\plotfiddle{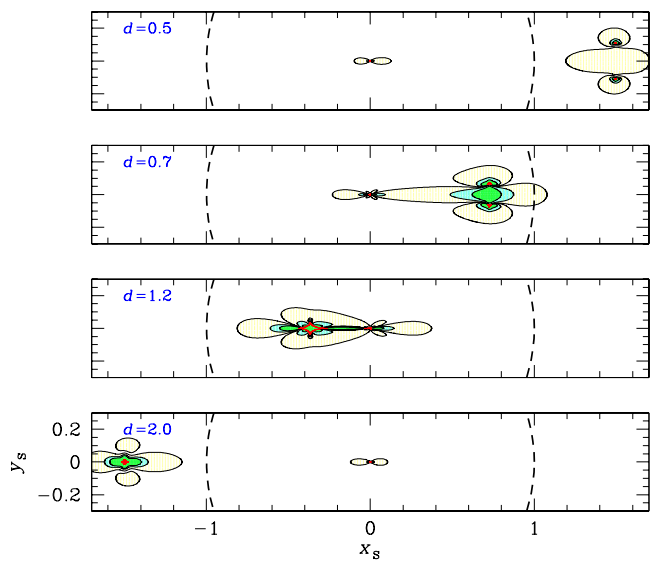}{0.0cm}{0}{48}{48}{-145}{-295}
\vskip7.0cm
\caption{The variation in the positions of the type I and II deviation regions.
Notations are same as those in the upper panel of Figure 1.}
\end{figure}

Due to the difference in the characteristics of the central and planetary 
caustics and the resulting anomalies, detecting planets by using the type II 
strategy has both advantages and disadvantages.  The greatest advantage of 
the type II strategy is that the time of anomalies, which occurs near the 
peak amplification, can be predicted in advance, and thus high time-resolution 
monitoring of the event is possible \citep{griest98}.  In addition, more 
accurate photometry can be performed because more photons will be available 
for high amplification events.  On the other hand, the disadvantage is that 
type II anomalies occur with a relatively low frequency due to the smaller 
size of the central caustic.  Then, the question is whether the planet 
detection rate under the type II strategy can be high enough to be comparable 
to the rate under the type I strategy.  In this section, we answer to this 
question by estimating the relative planet detection rates expected under
the two different types of planet search strategies.

To estimate the detection rates, we assume the following observational 
conditions and detection criteria.  For the type I strategy, all events with 
$A\gtrsim A_{\rm th}=1.34$ (i.e.\ $u\leq u_{\rm th}=1.0$) are assumed to be 
monitored in a round-the-clock manner during 8 hours per night on average.
Here $u_{\rm th}$ and $A_{\rm th}$ represent the threshold lens-source 
separation and the corresponding threshold amplification, which are required 
for the event to become the target of followup observations.  On the other 
hand, if the events are suspected to have high amplifications with $A \geq 
A_{\rm th}=10$ (i.e.\ $u\lesssim u_{\rm th} =0.1$), they become targets 
for intensive montoring by the type II strategy.  We assume that the type 
I strategy is employed during $-0.4 t_{\rm E} \leq t_{\rm obs} \leq 2.0 
t_{\rm E}$, while the type II strategy is employed only during the time of 
high amplifications with $A\geq A_{\rm th}$.  For planet detection, it is 
assumed that one should detect light curve deviations greater than a 
threshold value, $\epsilon_{\rm th}$, at least more than 5 times.  The 
adopted threshold deviations are $\epsilon_{\rm th} =5\%$ for the type I 
and  $1\%$ for the type II strategies, respectively.  We adopt the lower 
value of $\epsilon_{\rm th}$ for the type II strategy because the events 
monitored by this strategy will have higher amplifications.  For more 
sophisticated computations of the planet detection probability including 
not only more refined models of photometric precision but also the effects 
of finite source size and blending, see \citet{gaudi00}.  According to 
the current microlensing detection rate, there are on average several 
dozens of on-going events.  Then, considering the average total observation 
time of $\sim 8$ hours per night and $\sim 5$ -- 10 minutes of required 
time per event, one can observe on average $\sim 2$ -- 3 times for each 
event under the type I strategy.  However, since the frequency can be 
increased by employing multiple telescopes like the current followup 
observation teams \citep{albrow98, rhie99}, we test various monitoring 
frequencies per night of $f_{\rm obs}=2$, 5, and 10 for the type I strategy.  
For the type II strategy, we test $f_{\rm obs}=10$, 20, and 50.

In Figure 2, we present the distributions of the planet detection probability 
expected under the individual strategies with various monitoring frequencies 
as a function of the planet-lens separation for events caused by a planetary 
lens system with $q=10^{-3}$.  The events are assumed to have a common 
Einstein time scale of $t_{\rm E}= 20\ {\rm days}$, which corresponds to that 
of the bulge self-lensing event with a lens mass $1\ M_\odot$.  The region 
enclosed by the dot-dashed lines represents the lensing zone.  Note that our 
probability under the type II strategy are normalized by the total number of 
events with $A\geq 1.34$ unlike those of \citet{griest98}, who normalized by 
the number of only the intensively monitored high amplification events.  Since 
we applied $u_{\rm th}=0.1$, our probability is equivalent to approximately 
one tenth of their value.  In Table 1, we list the fractions of planet 
detection rate by the type II strategy out of the total rate (type I + type 
II) for various combinations of $f_{\rm obs}$ of the individual strategies.  
We estimate two sets of fractions.  The first set is for all planetary systems 
with separations $d\leq 5.0$ and the second set is for those with planets 
located within the lensing zone only.

The findings from Figure 2 and Table 1 are summarized as follows:
\begin{enumerate}
\item
As pointed out by \citet{griest98}, the planet detection efficiency per event, 
which is ten times of the probability presented in Figure 2, expected under
the type II strategy is very high.  We note, however, that our estimate is 
somewhat lower than their estimate.  We suspect the difference is caused 
because they assume that the type II monitoring strategy is applied during 
the entire duration of the event, while we assume that intensive monitoring 
is performed only during the time of high amplifications.  We actually obtain 
similar efficiency distributions to theirs under the observational conditions 
they assumed.

\item
In addition to the higher efficiency, the planet detection rate under the type 
II strategy is expected to comprise a substantial fraction of the total rate.
We find that the fraction will be $\sim 1/4$ -- 1/2 depending on the 
combinations of the individual strategies' monitoring frequencies.

\item
Compared to the type I strategy, the type II strategy is more efficient in 
detecting planets located outside of the lensing zone.  This is because while 
the planetary caustic lies at the position with increasing distance from the 
primary lens as the planet-lens separation becomes smaller or large than 
$d\sim 1.0$ \citep{wambsganss97}, the central caustic is located always near 
the primary lens regardless of the planet-lens separation (see Figure 3).  
As a result, while the type II efficiency decreases only by the decrease of 
the (central) caustic size, the type I efficiency is decreased additionally 
by the increasing distance of the (planetary) caustic location from the 
primary lens.

\item
The optimal monitoring frequency of the type II strategy will be $f_{\rm obs}
\sim 20$.  This is because observations with frequencies higher than this 
value will result in very slight increase in the planet detection rate (see
Table 1).  This optimal frequency can be easily achieved by the current 
followup programs even with a single telescope.

\end{enumerate}

\section{Conclusion}

We compare the planet detection rates expected by the type I and II followup 
observation strategies.  From this comparison, we find that the rate under 
the type II strategy will be substantial, comprising $\sim 1/4$ -- 1/2 of 
the total rate.  It is also found that compared to the type I strategy, the 
type II strategy is more efficient in detecting planets located outside of 
the lensing zone.  Therefore, in addition to yielding high planet detection 
efficiency for each high amplification event, employment of the type II
strategy will allow one not only to increase the total planet detection rate 
but also to probe the existence of planets with a wider range of separations.
Despite these benefits, implementation of the strategy requires only a modest 
improvement in the monitoring frequencies.

\acknowledgements
This work was supported by the grant 
from Korea Science \& Engineering Foundation (KOSEF).

\end{document}